# *First Principles Calculation of Topological Invariants of non-Hermitian Photonic Crystals*

**Filipa R. Prudêncio[1,2*], and Mário G. Silveirinha[1]**

[1]University of Lisbon – Instituto Superior Técnico and Instituto de Telecomunicações, Avenida Rovisco Pais 1, 1049-001 Lisbon, Portugal

[2]Instituto Universitário de Lisboa (ISCTE-IUL), Avenida das Forças Armadas 376, 1600-077 Lisbon, Portugal

## Abstract

The Chern topological numbers of a material platform are usually written in terms of the Berry curvature, which depends on the normal modes of the system. Here, we use a gauge invariant Green's function method to determine from "first principles" the topological invariants of photonic crystals. The proposed formalism does not require the calculation of the photonic band-structure, and can be easily implemented using the operators obtained with a standard plane-wave expansion. Furthermore, it is shown that the theory can be readily applied to the classification of topological phases of non-Hermitian photonic crystals with lossy or gainy materials, e.g., parity-time symmetric photonic crystals.

---

[*] Corresponding author: filipa.prudencio@lx.it.pt

# I. Introduction

Topological systems have fascinating and intriguing properties that can lead to unique physical effects and phenomena [1-16]. Topological methods became part of mainstream condensed matter physics with the discovery of the quantum Hall effect [1]. About one decade ago, the research of topological systems was extended to photonics [6, 7, 13, 14], and to other fields. Topological systems open a myriad of exciting opportunities in optics as they may allow for the propagation of light with suppressed back-scattering [8, 10]. Recently, it was shown that the photonic Chern number can be understood as the quantum of the fluctuation-induced light angular-momentum in a topological material cavity [15, 16, 17].

The topological properties of a physical system are usually linked to the spectrum of a two-parameter family of linear operators $\hat{H}_\mathbf{k}$. Typically, the spectrum is formed by the system natural modes (in condensed matter systems $\hat{H}_\mathbf{k}$ is the Hamiltonian of Bloch electronic states). It was recently discovered that the topological classification remains feasible even when the operator $\hat{H}_\mathbf{k}$ is non-Hermitian [18-25]; thereby lossy or gainy photonic systems are characterized by different topological phases.

In this article, we focus on Chern-type topological insulators. A non-trivial Chern phase can only occur when the system has a broken time reversal symmetry. Hence, in optics, non-trivial Chern insulators are necessarily nonreciprocal [26]. Each topological phase is characterized by an integer number (the Chern number), which is a topological invariant insensitive to weak perturbations of the Hamiltonian.

The Chern numbers of a (Hermitian) material system are traditionally obtained from the Berry curvature $\mathcal{F}_\mathbf{k}$ [13, 14, 27]. From the topological band theory, the gap Chern number is

$$\mathcal{C}_{\text{gap}} = \frac{1}{2\pi} \iint_{B.Z.} d^2\mathbf{k}\, \mathcal{F}_\mathbf{k} . \tag{1a}$$

The integral is over the first Brillouin zone (BZ) and $\mathcal{F}_\mathbf{k} = \sum_{n \in F} \mathcal{F}_{n\mathbf{k}}$ is the Berry curvature; the summation in $n$ is over all the "filled" photonic bands ($F$) below the gap, i.e., modes with $\omega_{n\mathbf{k}} < \omega_{\text{gap}}$, with $\omega_{\text{gap}}$ some frequency in the band gap. The Berry curvature of the $n$-th band ($\mathcal{F}_{n\mathbf{k}}$) is written in terms of the system eigenmodes ($\hat{H}_\mathbf{k} |n\mathbf{k}\rangle = \omega_{n\mathbf{k}} |n\mathbf{k}\rangle$) as [27, 28, 29]:

$$\mathcal{F}_{n\mathbf{k}} = \sum_{m\mathbf{k} \neq n\mathbf{k}} \frac{i}{(\omega_{n\mathbf{k}} - \omega_{m\mathbf{k}})^2} \left[ \langle n\mathbf{k} | \partial_1 \hat{H}_\mathbf{k} | m\mathbf{k} \rangle \langle m\mathbf{k} | \partial_2 \hat{H}_\mathbf{k} | n\mathbf{k} \rangle - 1 \leftrightarrow 2 \right] \tag{1b}$$

where $\partial_i = \partial / \partial k_i$ ($i$=1,2) with $k_1 = k_x$ and $k_2 = k_y$. The term $1 \leftrightarrow 2$ is obtained from the first term enclosed in rectangular brackets by exchanging the indices 1 and 2. Thus, from a computational point of view, the numerical calculation of the Chern invariants is a rather formidable problem: it generally requires finding the photonic band structure and all the Bloch states in the Brillouin zone. Moreover, Eq. (1b) is ill-defined due to a 0/0 removable singularity when the Bloch eigenstates are degenerate, i.e., when there are band crossings.

The situation is even more complex for non-Hermitian systems. In such a case the Chern number is found from a bi-orthogonal basis of left and right eigenstates of the non-Hermitian operator [19], i.e., from the eigenstates of $\hat{H}_\mathbf{k}$ and of $(\hat{H}_\mathbf{k})^\dagger$. The spectra of

$\hat{H}_\mathbf{k}$ and of $\left(\hat{H}_\mathbf{k}\right)^\dagger$ are not independent; their calculation requires the diagonalization of a non-Hermitian matrix which may be computationally demanding. The problem is especially complex for periodic structures, e.g., lossy nonreciprocal photonic crystals, where the relevant operator $\hat{H}_\mathbf{k}$ is isomorphous to a non-Hermitian matrix of infinite dimension.

Due to these difficulties, typically the topological invariants of photonic systems are determined using simplified models, e.g., relying on a tight-binding approximation, which allow for analytical developments. Even though justified in many cases, one needs to be careful which approximate methods, as the topological invariant is a *global* property of the physical system, while a tight-binding description does not always capture the *global* wave dynamics. Indeed, tight-binding models are often accurate only for a restricted range of wave vectors, rather than in the entire Brillouin zone. From a geometrical point of view the situation can be pictured as follows. Consider a spherical surface and another object almost coincident with the spherical surface but with a hole pierced in it, let us say the same sphere but with a cylindrical hole (with arbitrarily small, but finite, radius) that connects the north and south poles of the sphere. Even though locally (away from the two poles) the two objects are indistinguishable, they are topologically different, as they have a different number of holes. Thus, the topology of a physical system can be coincident with the topology of an approximate model only when the approximate model captures faithfully the wave dynamics in the *entire* parametric space, i.e., in the entire Brillouin zone.

In this article, we tackle the problem of first principles calculation, i.e., without using tight binding or other approximations, of the Chern number of non-Hermitian photonic systems. Our approach is based on Refs. [24, 28], where it was shown that the gap Chern numbers of non-Hermitian systems can be written in terms of the system Green's function. Different from the standard topological band theory, the Green's function approach is gauge invariant and does not require any detailed knowledge of the band structure or of the Bloch eigenstates. The method applies to both fermionic (see also Refs. [30-32]) and bosonic platforms (even in case of material dispersion). Different from topological band theory, it is unnecessary to compute the Chern invariants of the individual bands to find the gap topological invariant. The gap Chern number is directly obtained from an integral of the photonic Green's function over a contour in the complex frequency plane that links $-i\infty$ to $+i\infty$ and contained in the relevant band-gap.

The described theory was applied to electromagnetic continua in Refs. [24, 28]. Here, we tackle the more challenging and interesting case of photonic crystals. We show how by using the operators obtained from the well-known plane wave method [34] it is possible to find in a relatively simple and computationally inexpensive way the gap Chern number of topological photonic platforms. Furthermore, we study the impact of material loss on the topological invariants.

The article is organized as follows. In Sect. II we present a brief overview of the general Green's function method. In Sect. III we use the Green's function approach to determine the topological phases of lossless, lossy, and lossy-gainy magnetic-gyrotropic photonic crystals. A short summary of the key results is given in Sect. IV.

## II. Photonic Green's function formalism

In this Section, we briefly review the general Green's function formalism introduced in Refs. [24, 28] to calculate the Chern invariants of photonic platforms. The starting point is the generalized eigenvalue problem

$$\hat{L}_{\mathbf{k}} \cdot \mathbf{Q}_{n\mathbf{k}} = \omega_{n\mathbf{k}} \mathbf{M}_g \cdot \mathbf{Q}_{n\mathbf{k}}, \qquad (n=1,2,\ldots) \qquad (2)$$

with $\hat{L}_{\mathbf{k}}$ a generic differential operator and $\mathbf{M}_g$ a multiplication (matrix) operator. The operator $\hat{L}_{\mathbf{k}}$ is parameterized by the real wave vector $\mathbf{k} = k_x \hat{\mathbf{x}} + k_y \hat{\mathbf{y}}$. The operator $\mathbf{M}_g$ is independent of $\mathbf{k}$. Here, $\mathbf{Q}_{n\mathbf{k}}$ are the generalized eigenstates of $\hat{L}_{\mathbf{k}}$ and $\omega_{n\mathbf{k}}$ are the generalized eigenvalues. The objective is to determine the topological phases of $\hat{L}_{\mathbf{k}}$, or equivalently the topological phases of $\hat{H}_{\mathbf{k}} = \mathbf{M}_g^{-1} \cdot \hat{L}_{\mathbf{k}}$.

To this end, we introduce the system Green's function $\mathcal{G}_{\mathbf{k}}$, defined by

$$\mathcal{G}_{\mathbf{k}}(\omega) = i\left(\hat{L}_{\mathbf{k}} - \mathbf{M}_g \omega\right)^{-1}. \qquad (3)$$

The Green's function has poles at the eigenfrequencies $\omega = \omega_{n\mathbf{k}}$, but otherwise is an analytic function of frequency. Let us first consider, without loss of generality, that $\hat{L}_{\mathbf{k}}$ and $\mathbf{M}_g$ are Hermitian operators. In that case, the eigenfrequencies $\omega_{n\mathbf{k}}$ are real-valued numbers. Hence, the projection of the system band structure into the complex-frequency plane $\omega = \omega' + i\xi$ consists of line segments contained in the real-frequency axis (see Fig. 1a). The band gaps are the regions of the complex frequency plane that separate disconnected sets of eigenfrequencies. For example, with reference to Fig. 1a the region $\omega_L < \omega' < \omega_U$ is a band-gap (vertical strip shaded in yellow in Fig. 1a), as it separates two sets eigenfrequencies, i.e., two bands. This band-gap definition can be readily extended to non-Hermitian systems, with the difference that for non-Hermitian platforms the projected band structure is not restricted to the real-frequency axis. Hence, in the non-Hermitian case the projected band-structure can populate parts of the lower-half (for lossy systems) or upper-half (for gainy systems) complex-frequency plane [24]. In general, the band-gaps are vertical strips in the complex plane, i.e., of the form $\omega_L < \omega' < \omega_U$, where the Green's function is analytic (the vertical strip does not need to be rectangular and can have an arbitrary shape provided the initial and end points have $\xi = \mp\infty$ respectively).

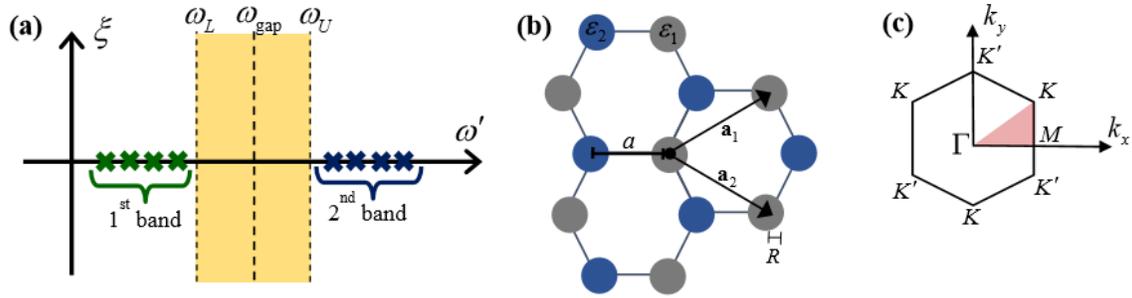

**Fig. 1 (a)** Illustration of the projection of the band structure of a Hermitian operator in the complex frequency plane $\omega = \omega' + i\xi$. The Green's function is an analytical function of frequency $\omega$ in the band gap represented by the vertical yellow strip. **(b)** Hexagonal array of magnetic-gyrotropic rods. The distance between nearest neighbors is $a$. **(c)** First Brillouin zone of the 2D lattice.

Each band-gap is associated with a topological invariant, the gap Chern number, given by [24, 28]:

$$\mathcal{C}_{\text{gap}} = \frac{1}{(2\pi)^2} \iint_{B.Z.} d^2\mathbf{k} \int_{\omega_{\text{gap}}-i\infty}^{\omega_{\text{gap}}+i\infty} d\omega \, \text{Tr}\{\partial_1 \mathcal{G}_{\mathbf{k}}^{-1} \cdot \mathcal{G}_{\mathbf{k}} \cdot \partial_2 \mathcal{G}_{\mathbf{k}}^{-1} \cdot \partial_\omega \mathcal{G}_{\mathbf{k}}\}, \qquad (4)$$

where $\text{Tr}\{...\}$ stands for the trace operator, $\partial_\omega = \partial/\partial\omega$ and $\partial_j \mathcal{G}_\mathbf{k}^{-1} = \partial \mathcal{G}_\mathbf{k}^{-1}/\partial k_j$ ($j$=1,2) with $k_1 = k_x$ and $k_2 = k_y$. The integral in $\omega$ is over a contour completely contained in the band-gap that joins the points $-i\infty$ and $+i\infty$. For simplicity, throughout the article it is assumed that the contour is a straight line of the form $\text{Re}\{\omega\} = \omega_{\text{gap}}$ with $\omega_{\text{gap}}$ some (real-valued) frequency in the gap (see Fig. 1a).

The derivatives in frequency and wave vector can be explicitly evaluated as $\partial_\omega \mathcal{G}_\mathbf{k} = -i\mathcal{G}_\mathbf{k} \cdot \mathbf{M}_g \cdot \mathcal{G}_\mathbf{k}$ and $\partial_j \mathcal{G}_\mathbf{k}^{-1} = -i\partial \hat{L}_\mathbf{k}/\partial k_j$ [24, 28]. Hence, the gap Chern number can be expressed as:

$$\mathcal{C}_{\text{gap}} = \frac{i}{(2\pi)^2} \iint_{B.Z.} d^2\mathbf{k} \int_{\omega_{\text{gap}}-i\infty}^{\omega_{\text{gap}}+i\infty} d\omega \, \text{Tr}\{\partial_1 \hat{L}_\mathbf{k} \cdot \mathcal{G}_\mathbf{k} \cdot \partial_2 \hat{L}_\mathbf{k} \cdot \mathcal{G}_\mathbf{k} \cdot \mathbf{M}_g \cdot \mathcal{G}_\mathbf{k}\}. \tag{5}$$

In order to numerically calculate the integral it is convenient to use the coordinates $\omega = \omega_{\text{gap}} + i\xi$ and $\mathbf{k} = \beta_1 \mathbf{b}_1 + \beta_2 \mathbf{b}_2$, where $\mathbf{b}_j$ are the reciprocal lattice primitive vectors of the photonic crystal and $-1/2 \leq \beta_j \leq 1/2$ ($j$=1,2). With these coordinate transformations we finally get:

$$\mathcal{C}_{\text{gap}} = \int_{-1/2}^{1/2}\int_{-1/2}^{1/2} d\beta_1 d\beta_2 \int_0^\infty d\xi \, g(\xi, \beta_1, \beta_2), \tag{6a}$$

$$g(\xi, \beta_1, \beta_2) = \frac{-1}{(2\pi)^2} |\mathbf{b}_1 \times \mathbf{b}_2|$$
$$\times \left[ \text{Tr}\{\partial_1 \hat{L}_\mathbf{k} \cdot \mathcal{G}_\mathbf{k} \cdot \partial_2 \hat{L}_\mathbf{k} \cdot \mathcal{G}_\mathbf{k} \cdot \mathbf{M}_g \cdot \mathcal{G}_\mathbf{k}\}\Big|_{\substack{\omega=\omega_{\text{gap}}+i\xi \\ \mathbf{k}=\beta_1\mathbf{b}_1+\beta_2\mathbf{b}_2}} + \text{Tr}\{\partial_1 \hat{L}_\mathbf{k} \cdot \mathcal{G}_\mathbf{k} \cdot \partial_2 \hat{L}_\mathbf{k} \cdot \mathcal{G}_\mathbf{k} \cdot \mathbf{M}_g \cdot \mathcal{G}_\mathbf{k}\}\Big|_{\substack{\omega=\omega_{\text{gap}}-i\xi \\ \mathbf{k}=\beta_1\mathbf{b}_1+\beta_2\mathbf{b}_2}} \right].$$

(6b)

In practice, the upper-limit of the integral in $\xi$ needs to be truncated: $\int_0^\infty d\xi \to \int_0^{\xi_{max}} d\xi$, where $\xi_{max}$ should be on the order of $c/a$ with $c$ the speed of light and $a$ the lattice constant. Typically, $g$ decays exponentially fast with $\xi$ and hence the integration in $\xi$ is quite efficient [24, 28]. In practice, the integrals in $\beta_1, \beta_2, \xi$ are done using numerical quadrature, e.g., using the trapezoidal or the Simpson rules.

## III. Magnetic-gyrotropic photonic crystal

### A. Physical model

To illustrate the ideas, we consider a photonic crystal formed by a hexagonal array of cylindrical rods with radius $R$ embedded in air as illustrated in Fig. 1b. The periodic structure contains two rods per unit cell, i.e., it is formed by two sub-lattices (honeycomb lattice). The direct lattice primitive vectors are taken equal to:

$$\mathbf{a}_1 = \frac{a}{2}\left(3\hat{\mathbf{x}} - \sqrt{3}\hat{\mathbf{y}}\right), \qquad \mathbf{a}_2 = \frac{a}{2}\left(3\hat{\mathbf{x}} + \sqrt{3}\hat{\mathbf{y}}\right), \qquad (7)$$

where $a$ is the distance between nearest neighbors (circles with different colors in Fig. 1b). The relative permittivity and permeability tensors of the photonic crystal are of the form:

$$\bar{\varepsilon} = \varepsilon \mathbf{1}_{3\times 3}, \qquad \bar{\mu} = \begin{pmatrix} \mu & i\kappa & 0 \\ -i\kappa & \mu & 0 \\ 0 & 0 & 1 \end{pmatrix}, \qquad (8)$$

with $\varepsilon = \varepsilon(x, y)$, $\mu_{11} = \mu_{22} = \mu(x, y)$ and $\mu_{12} = -\mu_{21} = i\kappa(x, y)$. The non-reciprocity parameter $\kappa$ vanishes in the air region and is equal to $\kappa_i$ ($i$=1,2) in the $i$-th sub-lattice of the hexagonal array. Hence, the rods material response is nonreciprocal ($\bar{\mu} \neq \bar{\mu}^T$) and gyrotropic. This type of material response occurs in natural ferrimagnetic materials (e.g., ferrites) biased with a magnetic field directed along the *z*-direction [33]. The parameters $\varepsilon$ and $\mu$ are identical to $\varepsilon = \mu = 1$ in the air region and to $\varepsilon_i$ and $\mu_i$ in the *i*-th sub-lattice (*i*=1,2) of the crystal. For simplicity, here we neglect material dispersion so that $\mu_i$ and $\kappa_i$ are frequency independent. The method can be generalized to include the effects of material dispersion [24, 28], but since there are a few nontrivial technicalities we leave that study for future work.

We consider waves with transverse electric (TE) polarization ( $\mathbf{E} = E_z \hat{\mathbf{z}}$ ) and propagation in the *xoy* plane, so that $E_z = E_z(x, y)$. From the Maxwell equations, $\nabla \times \mathbf{E} = i\omega\mu_0 \bar{\mu} \cdot \mathbf{H}$ and $\nabla \times \mathbf{H} = -i\omega\varepsilon_0 \varepsilon \mathbf{E}$, it readily follows that:

$$-\partial_x \left( \mu_{ef}^{-1} \partial_x E_z - i\chi \partial_y E_z \right) - \partial_y \left( \mu_{ef}^{-1} \partial_y E_z + i\chi \partial_x E_z \right) = \left( \frac{\omega}{c} \right)^2 \varepsilon E_z. \tag{9}$$

with $\mu_{ef} = \left( \mu^2 - \kappa^2 \right)/\mu$ and $\chi = \kappa/\left( \mu^2 - \kappa^2 \right)$. The secular equation can be written in the form:

$$\hat{L}(-i\nabla) \cdot E_z = \mathcal{E} \, \mathbf{M}_g \cdot E_z \tag{10}$$

with $\mathcal{E} = (\omega/c)^2$ and,

$$\begin{aligned} \mathbf{M}_g \cdot E_z &\equiv \varepsilon E_z \\ \hat{L} \cdot E_z &\equiv -\partial_x \left( \mu_{ef}^{-1} \partial_x E_z - i\chi \partial_y E_z \right) - \partial_y \left( \mu_{ef}^{-1} \partial_y E_z + i\chi \partial_x E_z \right) \end{aligned}. \tag{11}$$

As seen, $\mathbf{M}_g$ is a multiplication operator (multiplication by the material permittivity) and $\hat{L} = \hat{L}(-i\nabla)$ is a differential operator.

The Bloch modes associated with the wave vector $\mathbf{k} = k_x \hat{\mathbf{x}} + k_y \hat{\mathbf{y}}$ are of the form $E_z = e_z(x, y) e^{i\mathbf{k} \cdot \mathbf{r}}$, with the envelope $e_z(x, y)$ being a periodic function that satisfies the generalized eigenvalue problem:

$$\hat{L}_{\mathbf{k}} \cdot e_z = \mathcal{E} \mathbf{M}_g \cdot e_z, \qquad \text{with } \hat{L}_{\mathbf{k}} \equiv \hat{L}(-i\nabla + \mathbf{k}). \tag{12}$$

The operator $\hat{L}_{\mathbf{k}}$ is obtained from $\hat{L}$ with the substitutions $\partial_j \to \partial_j + ik_j$ ($j=x,y$). Evidently, the generalized eigenvalue problem is of the same type as in Eq. (2), and hence the topological phases of $\hat{L}_{\mathbf{k}}$ can be found using the formalism of Sect. II. Note that here the eigenvalues are $\mathcal{E} = (\omega/c)^2$, i.e., they are the *squared* eigenfrequencies of the photonic crystal (apart from a normalization factor). Thus, the gap Chern number can be found using Eq. (6) with the substitution $\omega \to \mathcal{E}$. For example, the Green's function must be defined as $\mathcal{G}_{\mathbf{k}}(\mathcal{E}) = i(\hat{L}_{\mathbf{k}} - \mathbf{M}_g \mathcal{E})^{-1}$ and the integration in $\xi$ is associated with the contour $\mathcal{E} = \mathcal{E}_{\text{gap}} + i\xi$, where $\mathcal{E}_{\text{gap}} = (\omega_{\text{gap}}/c)^2$.

### B. Band structure

The band structure of the photonic crystal can be found with the plane wave method [34]. To this end, the electric field envelope $e_z$ is expanded into plane waves as $e_z = \sum_{\mathbf{J}} c_{\mathbf{J}}^E e^{i\mathbf{G}_{\mathbf{J}} \cdot \mathbf{r}}$. Here, $\mathbf{G}_{\mathbf{J}} \equiv j_1 \mathbf{b}_1 + j_2 \mathbf{b}_2$ is a generic reciprocal lattice primitive vector, $\mathbf{J} = (j_1, j_2)$ represents a pair of integer numbers, and $\mathbf{b}_i$ are the reciprocal lattice primitive vectors defined by $\mathbf{a}_i \cdot \mathbf{b}_j = 2\pi \delta_{i,j}$ with $i, j = 1, 2$ [34].

In a plane wave basis, the operators $\mathbf{M}_g$ and $\hat{L}_{\mathbf{k}}$ are represented by matrices with infinite dimension. Explicit formulas for these matrices can be found in Appendix A [Eq.

(A6)]. In practice, the electric field plane wave expansion is truncated enforcing that $\mathbf{J} = (j_1, j_2)$ with $|j_i| \le j_{max}$ with $i=1,2$. In these circumstances, $\mathbf{M}_g$ and $\hat{L}_\mathbf{k}$ are given by matrices with dimension $(2j_{max}+1)^2 \times (2j_{max}+1)^2$ and Eq. (12) is reduced to a generalized matrix eigenvalue problem [Eq. (A7)].

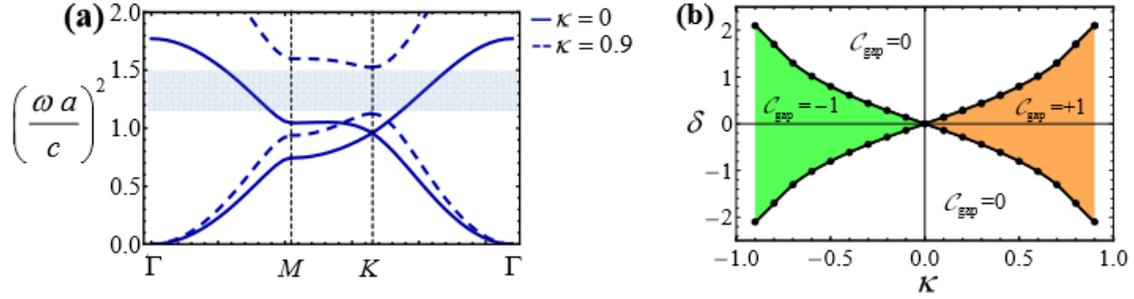

**Fig. 2** (a) Photonic band structure of a lossless gyrotropic photonic crystal; the material parameters of the cylindrical rods are $\varepsilon_1 = \varepsilon_2 = 12$ and $\mu_1 = \mu_2 = 1$. (i) Solid lines: reciprocal system ($\kappa = 0$). (ii) Dashed lines: nonreciprocal system ($\kappa = 0.9$). (b) Phase diagram of the first band gap of the photonic crystal showing the topological gap Chern number for different combinations of the nonreciprocity parameter ($\kappa$) and of the spatial-asymmetry parameter ($\delta$). The material parameters of the rods are $\varepsilon_1 = 12+\delta$, $\varepsilon_2 = 12-\delta$ and $\mu_1 = \mu_2 = 1$. The different topological phases are shaded with different colors. The boundaries between the different regions are obtained with an interpolation of the numerically calculated pairs $(\kappa, \delta)$ for which the band-gap closes (discrete black points).

The calculated band structure is plotted in Fig. 2a, for a reciprocal photonic crystal ($\kappa = 0$, solid lines) and for a nonreciprocal crystal ($\kappa = 0.9$, dashed lines). A sketch of the Brillouin zone and the definition of the relevant high-symmetry points can be found in Fig. 1c. The radius of the scattering centers is $R = 0.346a$ and the gyrotropic material constitutive parameters are taken equal to $\varepsilon_1 = \varepsilon_2 = 12$ and $\mu_1 = \mu_2 = 1$. As seen in Fig. 2a, for the reciprocal case, the bands touch at the Dirac point (K) due the symmetry of the hexagonal lattice ($\varepsilon_1 = \varepsilon_2$). Indeed, the reciprocal photonic crystal is a photonic analogue of graphene [35]. When a static magnetic field is applied to the system, so that $\kappa \neq 0$, the degeneracy around the Dirac points is lifted, leading to a complete photonic band-gap. For $\kappa = 0.9$, the band gap is determined by $1.12/a^2 < \mathcal{E} < 1.53/a^2$. The plane wave expansion was truncated with $|j_i| \leq j_{max} = 3$.

Analogous to the Haldane model [2, 12, 36], a band-gap between the first and second bands can be opened either by breaking the time-reversal symmetry ($\kappa \neq 0$, as illustrated in Fig. 2a) or, alternatively, by introducing some structural asymmetry between the two sub-lattices of the hexagonal array so that the inversion symmetry is broken. To model the latter situation, we introduce a spatial-asymmetry parameter ($\delta$) that controls the permittivity of the cylindrical rods as $\varepsilon_1 = 12 + \delta$ and $\varepsilon_2 = 12 - \delta$. For $\delta \neq 0$ the inversion symmetry of the system is broken. We numerically checked that for specific combinations of the parameters $(\kappa, \delta)$ the band-gap between the first and second bands is closed. These pairs of $(\kappa, \delta)$ are represented by the continuous black lines in Fig. 2b.

### C. Topological phases of a lossless system

The gap Chern number can be found by feeding the matrices that represent $\mathbf{M}_g$ and $\hat{L}_\mathbf{k}$ in the plane wave basis [Eq. (A6)] into the integral (6). Thus, the Chern number is calculated using the matrices of the standard plane wave method. The matrices $\partial_i \hat{L}_\mathbf{k}$ can be evaluated analytically [see Eq. (A8)]. Note that in the plane wave basis $\mathcal{G}_\mathbf{k}(\mathcal{E}) = i(\hat{L}_\mathbf{k} - \mathbf{M}_g \mathcal{E})^{-1}$ is a matrix. All the matrices have dimensions $(2j_{max} + 1)^2 \times (2j_{max} + 1)^2$.

In our numerical code, the integral (6) is evaluated using the trapezoidal rule. The Brillouin zone is uniformly sampled with $N_1 \times N_2$ points in the $(\beta_1, \beta_2)$ coordinates. The integral along the imaginary axis ($\mathcal{E} = \mathcal{E}_{gap} + i\xi$) is truncated at $\xi_{max} = 5/a^2$ and the integrand is sampled with $N_\xi$ points. The parameter $\mathcal{E}_{gap}$ is taken as the mid-point of the gap and we use $j_{max} = 3$.

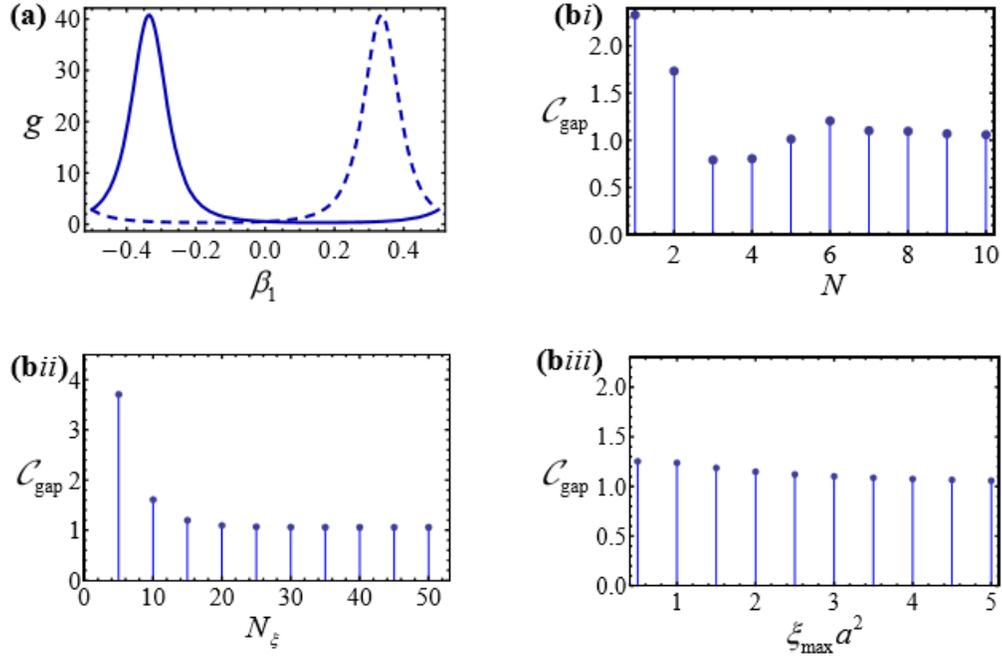

**Fig. 3** (a) Integrand of Chern number integral $g$ [Eq. (6b)] (in arbitrary units) as a function of $\beta_1$ for $\beta_2 = 1/3$ (solid line) and $\beta_2 = -1/3$ (dashed line). (b) The numerically calculated gap Chern number $C_{gap}$ as a function of (i) $N$. (ii) $N_\xi$. (iii) $\xi_{max}$. The parameters of the photonic crystal are the same as in Fig. 2a with $(\kappa, \delta) = (0.9, 0)$. The values of $N$, $N_\xi$, $\xi_{max}$ are 10, 50, and $5/a^2$, respectively, except for the parameter shown in the horizontal axis of a plot.

Figure 3a depicts the integrand of Eq. (6) as a function of $\beta_1$ for $\xi = 0$ and $\beta_2 = 1/3$ (solid line) and $\beta_2 = -1/3$ (dashed line). The photonic crystal parameters are the same as in Fig. 2a with $(\kappa, \delta) = (0.9, 0)$ and we take $\mathcal{E}_{gap} = 1.325/a^2$. As seen, the integrand is peaked near $\beta_1 = \mp 1/3$, which correspond to the coordinates of the $K'$ and $K$ points, respectively. This reveals that the topological charge is concentrated near the two Dirac points.

Figures 3b*i*-b*iii* show the numerically calculated gap Chern number as a function of $N \equiv N_1 = N_2$, $N_\xi$ and $\xi_{max}$, respectively. As seen, for modestly large values of $N$, $N_\xi$ and $\xi_{max}$, the numerical result converges to $\mathcal{C}_{gap} = 1$, consistent with the topological nature of the Chern number. The computation time of each Chern number is on the order of minutes in a standard personal computer with high-level language programming (Wolfram Mathematica).

The phase diagram of the ferrite photonic crystal is plotted in Fig. 2b. The diagram shows the gap Chern number for different combinations of the nonreciprocity and spatial-asymmetry parameters $(\kappa, \delta)$. The different topological phases are shaded with different colors and the corresponding gap Chern numbers are shown in the insets. As previously mentioned, the two black curves show the combinations of $(\kappa, \delta)$ for which the band gap closes. As seen, when the inversion symmetry breaking dominates (large values of $|\delta|$) the photonic crystal is topologically trivial. In contrast, when the time-reversal symmetry breaking dominates (large values of $|\kappa|$) the system is topologically nontrivial and is characterized by the gap Chern number $\mathcal{C}_{gap} = \text{sgn}(\kappa)$. According to the bulk-edge correspondence, the gap Chern number gives the net number of unidirectional edge states at the boundary of the photonic crystal for opaque-type boundary conditions (e.g., a perfect electric conducting boundary) [16, 37-41].

## D. Non-Hermitian systems

The formalism can be applied with no modifications to take into account the effect of material dissipation in the cylindrical rods. Non-energy conserving (non-Hermitian) platforms have recently raised a lot of interest due to the exotic physics of systems with exceptional points [42]. For simplicity, here we model the material loss by considering that $\mu$ is complex valued: $\mu = \mu' + i\mu''$. As in the previous subsection, we take $\mu' = 1$.

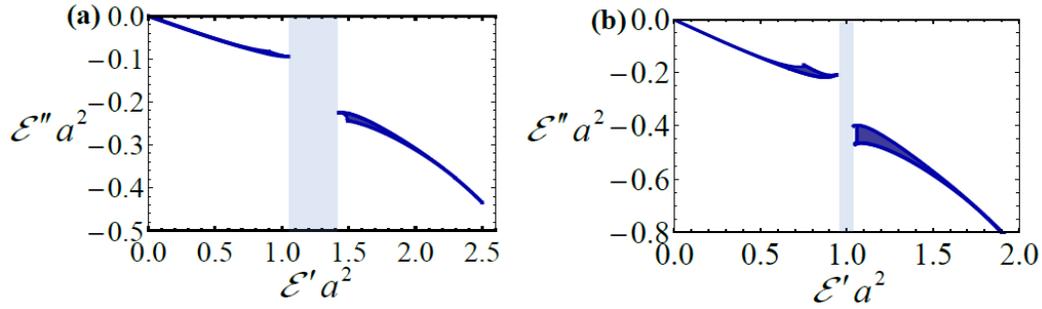

**Fig. 4** Band structure projected in the complex $\mathcal{E} = \mathcal{E}' + i\mathcal{E}''$ plane for a lossy magnetic-gyrotropic photonic crystal. The bands are projected onto the regions filled in dark-blue color. The photonic crystal parameters are as in Fig. 2a, except that $\mu_1 = \mu_2 = 1 + i\mu''$. The loss parameter is (**a**) $\mu'' = 0.1$ and (**b**) $\mu'' = 0.5$.

Figure 4 shows the band structure of the non-Hermitian photonic crystal projected on the $\mathcal{E}$-plane for two different values of loss parameter: $\mu'' = 0.1$ [Fig. 4a] and $\mu'' = 0.5$ [Fig. 4b]. The projected band structure represents the locus of $\mathcal{E} = \mathcal{E}' + i\mathcal{E}''$ as a function of the real-valued wave vector of the Bloch modes with $\mathcal{E} = (\omega a/c)^2$. For a lossy system the projected band structure lies in the lower-half frequency plane, different from the lossless case [Fig. 2a] where $\mathcal{E}$ is real-valued. As seen in Fig. 4, for both examples there is a vertical strip of the $\mathcal{E}$-plane (shaded light-blue region) free of natural modes. This vertical strip represents the band-gap. The two projected bands remain disconnected even in case of relatively strong material dissipation. In other words, typically the material dissipation does not close the band-gap.

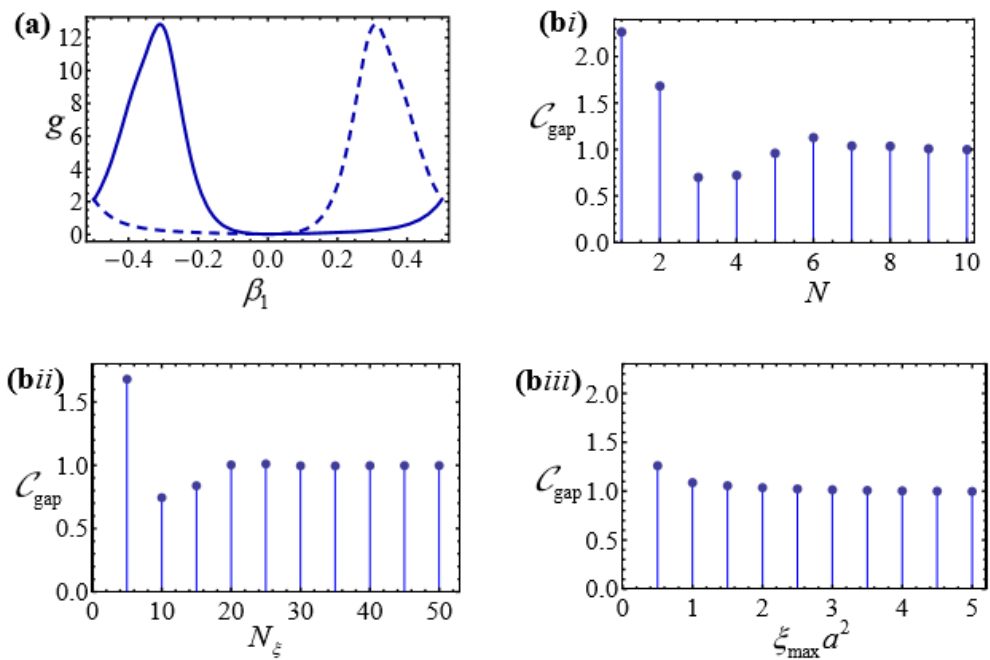

**Fig. 5** Study analogous to Fig. 3 but for a lossy magnetic-gyrotropic photonic crystal with $\mu_1 = \mu_2 = 1 + i\mu''$ with $\mu'' = 0.1$.

Figure 5 reports a study analogous to that of Fig. 3, but for a lossy photonic crystal with $\mu_1 = \mu_2 = 1 + i0.1$. As in the lossless case, the integrand $g$ is peaked at the Dirac points $K'$ and $K$ (see Fig. 5a), but now the topological charge is distributed more evenly through the wave vector space. The convergence rate and the computational effort to find the gap Chern number of the non-Hermitian system are similar to the lossless case (see Figs. 5b$i$-b$iii$). Furthermore, the numerical results confirm that the band gap of the projected band structure [Fig. 4a] is topologically non-trivial. The topological number is insensitive to the value of $\mu''$ in the range $0 \le \mu'' < 0.5$, as in this range the band-gap remains open. Thus, the topological properties of the photonic crystal are strongly robust to the dissipation effects.

The formalism can also be applied with no modifications to gainy systems. To illustrate this we consider a parity-time ($\mathcal{PT}$) symmetric [43, 44] gyrotropic photonic crystal. $\mathcal{PT}$-symmetric systems have rather unique features and can be implemented at optics through a judicious inclusion of gain/loss regions [45-48] or with moving media [49, 50]. The spectrum of $\mathcal{PT}$-symmetric systems is real-valued when the eigenfunctions simultaneously diagonalize the system Hamiltonian and the $\mathcal{PT}$ operator. Otherwise, the spectrum can be complex-valued, which corresponds to a spontaneously broken $\mathcal{PT}$-symmetry.

In our case, the $\mathcal{PT}$-symmetry can be enforced by assuming that $\mu_1 = 1 + i\mu''$ and $\mu_2 = 1 - i\mu''$ with $\mu'' > 0$, so that the first sub-lattice of cylinders is formed by dissipative elements and the second sub-lattice by gainy elements. The rest of the structural parameters are as in Fig. 2a.

Figure 6 shows the band structure of the $\mathcal{PT}$-symmetric gyrotropic photonic crystal projected on the $\mathcal{E}$-plane for four different cases: $\mu'' = 0.1$ [Fig. 6a], $\mu'' = 2$ [Fig. 6b], $\mu'' = 2.1$ [Fig. 6c] and $\mu'' = 2.2$ [Fig. 6d]. As seen in Fig. 6, for a lossy-gainy system the projected band structure lies both in the lower and upper half frequency planes, different from the lossy case [Fig. 4] where $\mathcal{E}''$ is always negative. Because of the $\mathcal{PT}$-symmetry the projected band structure exhibits a mirror-symmetry with respect to the real-frequency axis. Analogous to Fig. 4, the bands are projected onto the regions filled in dark-blue color. The boundary of the projected band structure corresponds roughly to the projection of the Brillouin zone path $\Gamma - M - K - \Gamma - M - K' - \Gamma$ [51]. For low values of

$\mu''$ [Fig. 6a], the first two projected bands are disconnected. The gap topological number is numerically evaluated in the same manner as in the previous examples and is equal to $C_{\text{gap}} = \text{sgn}(\kappa)$, consistent with Figs. 3 and 5. Therefore, as could be expected, moderate values of $\mu''$ do not affect the topological properties of the system. For larger values of the loss-gain parameter $\mu''$ [Fig. 6b], the band gap between the first two projected bands becomes very narrow. The band-gap closes approximately for $\mu'' = 2.1$, when the two bands touch at two distinct points [Fig. 6c]. For larger values of $\mu''$ [Fig. 6d], the two bands remain connected and the topological classification is not feasible.

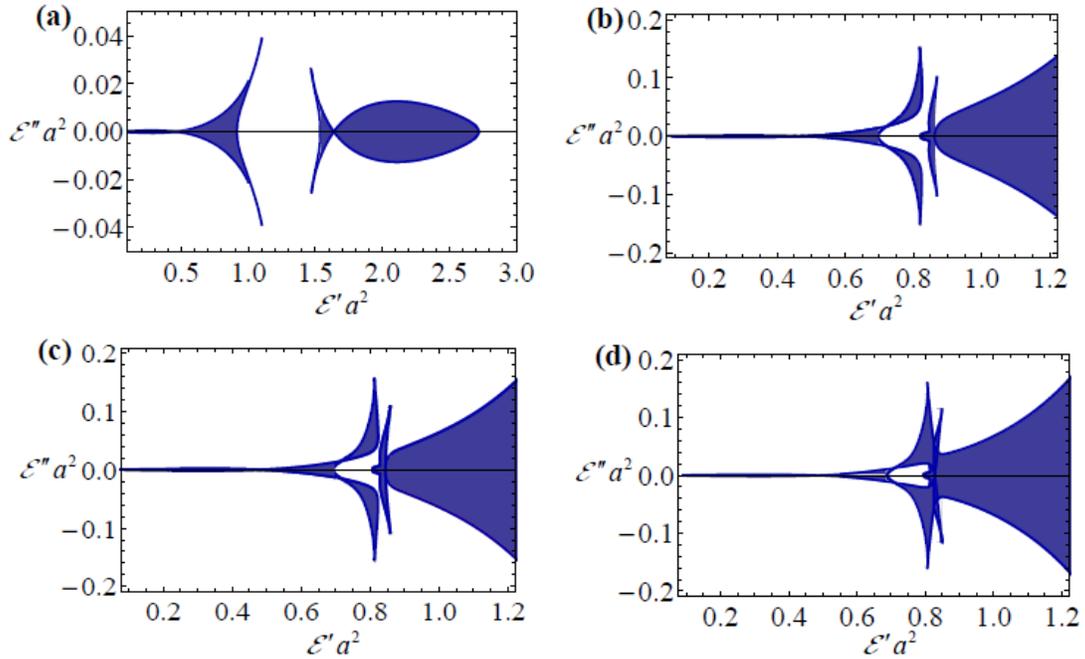

**Fig. 6** Band structure projected in the complex $\mathcal{E} = \mathcal{E}' + i\mathcal{E}''$ plane for a $\mathcal{PT}$ symmetric magnetic-gyrotropic photonic crystal. The photonic crystal parameters are as in Fig. 2a, except that $\mu_1 = 1 + i\mu''$ and $\mu_2 = 1 - i\mu''$. The loss-gain parameter $\mu''$ is (**a**) $\mu'' = 0.1$. (**b**) $\mu'' = 2$. (**c**) $\mu'' = 2.1$ (**d**) $\mu'' = 2.2$.

# IV. Summary

We used a Green's function method to calculate from "first principles" the topological invariants of Hermitian and non-Hermitian photonic crystals with a broken time-reversal symmetry. The main advantage of our formalism is that it does not require any detailed knowledge of the photonic band structure or of the Bloch modes. In particular, different from the topological band theory, the Green's function approach can be applied with no modifications when the different photonic bands cross at one or more points of the Brillouin zone. The computational effort for the non-Hermitian case is essentially the same as for the Hermitian case. The Green's function is numerically calculated using the standard plane-wave method. We applied the formalism to magnetic-gyrotropic photonic crystals. It was shown that the topological phases of a photonic crystal are strongly robust to non-Hermitian perturbations (dissipation and/or gain). We expect that our work will find widespread application in the characterization of emergent topological photonic platforms.

**Acknowledgements**

This work was supported by the IET under the A F Harvey Engineering Research Prize and by Fundação para Ciência e a Tecnologia (FCT) under project UIDB/50008/2020. F. R. Prudêncio acknowledges financial support by FCT under the Post-Doctoral fellowship SFRH/BPD/108823/2015.

# Appendix A: Plane wave representation of the operators $\mathbf{M}_g$ and $\hat{L}_\mathbf{k}$

In the following, we obtain the representations of the operators $\mathbf{M}_g$ and $\hat{L}_\mathbf{k}$ [see Eqs. (11)-(12)] in a plane wave basis [34].

To begin with, we expand the periodic functions $\mu_{ef}^{-1}$, $\chi$ and $\varepsilon$ into a Fourier series:

$$\mu_{ef}^{-1} = \sum_{\mathbf{J}} p_{\mu_{ef}^{-1},\mathbf{J}} e^{i\mathbf{G_J}\cdot\mathbf{r}}, \quad \chi = \sum_{\mathbf{J}} p_{\chi,\mathbf{J}} e^{i\mathbf{G_J}\cdot\mathbf{r}}, \quad \text{and} \quad \varepsilon = \sum_{\mathbf{J}} p_{\varepsilon,\mathbf{J}} e^{i\mathbf{G_J}\cdot\mathbf{r}}. \quad (A1)$$

The Fourier coefficients are $p_{\mu_{ef}^{-1},\mathbf{J}}$, $p_{\chi,\mathbf{J}}$, and $p_{\varepsilon,\mathbf{J}}$, respectively. For a generic function $g$ the Fourier coefficients are $p_{g,\mathbf{I}} = \frac{1}{A_{cell}} \int_{cell} g(\mathbf{r}) e^{-i\mathbf{G_I}\cdot\mathbf{r}} d^2\mathbf{r}$. For the geometry of Fig. 1b, the function $g$ (where $g$ can stand either for $\mu_{ef}^{-1}$, $\chi$ or $\varepsilon$) is sectionally constant. Let us suppose that $g = g_b$ in the air (background) region and $g = g_1$ and $g = g_2$ in the first and second sub-lattices of the hexagonal array, respectively. A straightforward calculation shows that [34]:

$$p_{g,\mathbf{I}} = g_b \delta_{\mathbf{I},0} + \sum_{i=1,2} f_{V,i} (g_i - g_b) e^{-i\mathbf{G_I}\cdot\mathbf{r}_{0,i}} \frac{2J_1(|\mathbf{G_I}|R_i)}{|\mathbf{G_I}|R_i}, \quad (A2)$$

where $\delta_{\mathbf{I},0}$ is Kronecker's symbol, $J_1$ is the cylindrical Bessel function of first kind and first order, $R_i$ is the radius of the rods of the $i$-th array, $\mathbf{r}_{0,i}$ gives the position of the $i$-th rod in the unit cell, and $f_{V,i} = \pi R_i^2 / A_{cell}$ with $A_{cell} = |\mathbf{b}_1 \times \mathbf{b}_2|$ the area of the unit cell. Note that $p_{g,0} = g_b + \sum_i f_{V,i} (g_i - g_b)$.

Consider now the operators defined in Eq. (11) with the electric field envelope ($e_z = E_z e^{-i\mathbf{k}\cdot\mathbf{r}}$) expanded in terms of plane waves as: $e_z = \sum_{\mathbf{J}} c_{\mathbf{J}}^E e^{i\mathbf{G_J}\cdot\mathbf{r}}$. Then, it is simple to check that:

$$\mu_{ef}^{-1} \partial_x E_z = \sum_{\mathbf{I},\mathbf{J}} e^{i(\mathbf{k}+\mathbf{G_I})\cdot\mathbf{r}} i\hat{\mathbf{x}} \cdot (\mathbf{k}+\mathbf{G_J}) p_{\mu_{ef}^{-1},\mathbf{I}-\mathbf{J}} c_{\mathbf{J}}^E. \quad (A3)$$

From here, it follows that:

$$-\partial_x \left( \mu_{ef}^{-1} \partial_x E_z \right) = \sum_{\mathbf{I},\mathbf{J}} e^{i(\mathbf{k}+\mathbf{G_I})\cdot\mathbf{r}} \hat{\mathbf{x}} \cdot (\mathbf{k}+\mathbf{G_J}) \hat{\mathbf{x}} \cdot (\mathbf{k}+\mathbf{G_I}) p_{\mu_{ef}^{-1},\mathbf{I}-\mathbf{J}} c_{\mathbf{J}}^E. \quad (A4)$$

Proceeding in the same way to calculate the other terms in Eq. (11), it is found that:

$$\hat{L} \cdot E_z = \sum_{\mathbf{I},\mathbf{J}} e^{i(\mathbf{k}+\mathbf{G}_\mathbf{I})\cdot\mathbf{r}} \left[ \hat{\mathbf{x}}\cdot(\mathbf{k}+\mathbf{G}_\mathbf{I})\hat{\mathbf{x}}\cdot(\mathbf{k}+\mathbf{G}_\mathbf{J}) + \hat{\mathbf{y}}\cdot(\mathbf{k}+\mathbf{G}_\mathbf{I})\hat{\mathbf{y}}\cdot(\mathbf{k}+\mathbf{G}_\mathbf{J}) \right] p_{\mu_{ef}^{-1},\mathbf{I}-\mathbf{J}} c_\mathbf{J}^E$$
$$+ \sum_{\mathbf{I},\mathbf{J}} e^{i(\mathbf{k}+\mathbf{G}_\mathbf{I})\cdot\mathbf{r}} i \left[ \hat{\mathbf{y}}\cdot(\mathbf{k}+\mathbf{G}_\mathbf{I})\hat{\mathbf{x}}\cdot(\mathbf{k}+\mathbf{G}_\mathbf{J}) - \hat{\mathbf{x}}\cdot(\mathbf{k}+\mathbf{G}_\mathbf{I})\hat{\mathbf{y}}\cdot(\mathbf{k}+\mathbf{G}_\mathbf{J}) \right] p_{\chi,\mathbf{I}-\mathbf{J}} c_\mathbf{J}^E \quad \text{(A5)}$$

Furthermore, it is clear that $\mathbf{M}_g \cdot E_z = \sum_{\mathbf{I},\mathbf{J}} e^{i(\mathbf{k}+\mathbf{G}_\mathbf{I})\cdot\mathbf{r}} p_{\varepsilon,\mathbf{I}-\mathbf{J}} c_\mathbf{J}^E$. Since $\hat{L}_\mathbf{k}\cdot e_z = e^{-i\mathbf{k}\cdot\mathbf{r}} \hat{L}\cdot E_z$, etc, the operators $\hat{L}_\mathbf{k}$ and $\mathbf{M}_g$ are represented by the matrices $\mathbf{M}_g \to [M_{\mathbf{I},\mathbf{J}}]$ and $\hat{L}_\mathbf{k} \to [L_{\mathbf{I},\mathbf{J}}]$ with:

$$\begin{aligned}
[M_{\mathbf{I},\mathbf{J}}] &= [p_{\varepsilon,\mathbf{I}-\mathbf{J}}] \\
[L_{\mathbf{I},\mathbf{J}}] &= \left[ (\mathbf{k}+\mathbf{G}_\mathbf{I})\cdot(\mathbf{k}+\mathbf{G}_\mathbf{J}) p_{\mu_{ef}^{-1},\mathbf{I}-\mathbf{J}} + i\left[(\mathbf{k}+\mathbf{G}_\mathbf{J})\times(\mathbf{k}+\mathbf{G}_\mathbf{I})\right]\cdot\hat{\mathbf{z}}\, p_{\chi,\mathbf{I}-\mathbf{J}} \right]
\end{aligned} \quad \text{(A6)}$$

Note that $[L_{\mathbf{I},\mathbf{J}}]$ ($[M_{\mathbf{I},\mathbf{J}}]$) stands for a matrix with element $L_{\mathbf{I},\mathbf{J}}$ ($M_{\mathbf{I},\mathbf{J}}$) in line $\mathbf{I}$ and column $\mathbf{J}$. The generalized eigensystem (12) is equivalent to:

$$[L_{\mathbf{I},\mathbf{J}}]\cdot[c_\mathbf{J}^E] = \mathcal{E}[M_{\mathbf{I},\mathbf{J}}]\cdot[c_\mathbf{J}^E]. \quad \text{(A7)}$$

Finally, we note the operators $\partial_i \hat{L}_\mathbf{k}$ are represented by

$$\partial_i \hat{L}_\mathbf{k} \to \left[ \hat{\mathbf{u}}_i \cdot (2\mathbf{k}+\mathbf{G}_\mathbf{I}+\mathbf{G}_\mathbf{J}) p_{\mu_{ef}^{-1},\mathbf{I}-\mathbf{J}} + i\left[\hat{\mathbf{u}}_i \times (\mathbf{G}_\mathbf{I}-\mathbf{G}_\mathbf{J})\right]\cdot\hat{\mathbf{z}}\, p_{\chi,\mathbf{I}-\mathbf{J}} \right] \quad \text{(A8)}$$

in the plane wave basis, where $\hat{\mathbf{u}}_1 = \hat{\mathbf{x}}$ and $\hat{\mathbf{u}}_2 = \hat{\mathbf{y}}$ are unit vectors along the coordinates axes.

[51] Figure 6 was obtained under the simplifying hypothesis that the boundary of the projected band structure is determined by the Brillouin zone path $\Gamma - M - K - \Gamma - M - K' - \Gamma$. We numerically checked that such approximation leads to results almost coincident with the exact projected band structure.